\documentclass[a4paper]{article}
\usepackage[a4paper,top=3cm,bottom=2cm,left=3cm,right=3cm,marginparwidth=1.75cm]{geometry}
\usepackage{amsfonts}
\usepackage{bm}
\usepackage{extarrows}
\usepackage{amssymb}
\usepackage{color}
\usepackage[all]{xy}
\usepackage{graphicx}
\usepackage{braket}
\usepackage{amsmath}
\usepackage{appendix}
\usepackage{graphicx}
\usepackage[colorinlistoftodos]{todonotes}
\usepackage{amsthm}
\usepackage{mathrsfs}
\usepackage{amssymb}
\usepackage{accents}
\usepackage{extarrows}
\usepackage{makecell}
\usepackage{authblk}
\usepackage{url}
\usepackage{hyperref}
\usepackage{cite}
\usepackage{eufrak}
\hypersetup{colorlinks,
            citecolor=black,
            linkcolor=black,
            urlcolor=green,
            pdftex}

\usepackage[english]{babel}
\usepackage[utf8x]{inputenc}
\usepackage[T1]{fontenc}

\usepackage[normalem]{ulem}

\title{Superposition type coherent states in all dimensional loop quantum gravity}
\author[1,2]{Gaoping Long \footnote{201731140005@mail.bnu.edu.cn}}
\author[3]{Cong Zhang\footnote{czhang@fuw.edu.pl}}
\author[1]{Xiangdong Zhang \footnote{scxdzhang@scut.edu.cn}\thanks{corresponding author}}

\affil[1]{Department of Physics, South China University of Technology, Guangzhou 510641, China}
\affil[2]{Department of Physics, Beijing Normal University, Beijing 100875, China}
\affil[3]{Faculty of Physics, University of Warsaw, Pasteura 5, 02-093 Warsaw, Poland}

\date{}

\begin{document}

\maketitle

\begin{abstract}
We propose a new kind of coherent state for the general $SO(D+1)$
formulation of loop quantum gravity in the $(1+D)$-dimensional space-time. Instead of Thiemann's coherent state for $SO(D+1)$ gauge theory, our coherent spin-network state is given by constructing proper superposition over quantum numbers of the spin-networks with vertices labelled by the coherent intertwiners. Such superposition type coherent states are labelled by the so-called generalized twisted geometric variables which capture the geometric meaning of discretized general relativity. We study the basic properties of this kind of coherent states, i.e., the completeness and peakedness property. Moreover, we show that the superposition type coherent states are consistent with Thiemann's coherent state for $SO(D+1)$ gauge theory in large $\eta$ limit.

\end{abstract}

\section{Introduction}
Coherent states in loop quantum gravity (LQG) , as a kind of quantum state which most closely resembles the classical geometry,  plays an important role for the exploration of many directions. In (1+3)-dimensional standard LQG, construction of the Hilbert space relies on graphs $\gamma$ in the spatial manifold \cite{Ashtekar2012Background,Han2005FUNDAMENTAL,thiemann2007modern,rovelli2007quantum,RovelliBook2}. Given a graph $\gamma$ consisting of $|E(\gamma)|$ edges, then the mathematical structure of LQG associated to $\gamma$ is the same as that for the quantum mechanics of ``free'' particle on the group manifold $SU(2)^{|E(\gamma)|}$. Therefore, the Hall's heat-kernels coherent state for compact Lie group can be borrowed to construct a type of coherent state in the standard LQG theory which is referred to as the Thiemann's coherent state \cite{ThiemannComplexifierCoherentStates,Thomas2001Gauge,2001Gauge,2000Gauge}.  The properties of the Thiemann's coherent states have been fully studied and widely used \cite{Han_2020,Han_2020semiclassical,zhang2021firstorder}. Particularly, it was shown that these coherent state possess a well-behaved peakedness property in the discrete phase space of LQG, and their ``Ehrenfest property''  was also shown to guarantee the coincidence between the expectation values of the elementary operators $\hat O$ and the evaluation of $O$ on the phase space.
Apart from the Thiemann's coherent states, another type of coherent states in $SU(2)$ LQG, introduced and applied to analyze the asymptotics of the EPRL spin foam model in \cite{Rovelli_2006,Bianchi_2009}, studied further in \cite{Bianchi_2010,Calcinari_2020}, takes simpler formulation and more obvious geometrical meanings. Such kind of coherent state is given by constructing proper superposition over spins of the spin-networks with vertices labelled by the coherent intertwiners \cite{Livine:2007Nsfv}. We will refer to these coherent state as the superposition type coherent state in the current work. It has been shown that the superposition type coherent states can be derived from the hear-kernel coherent states (or Thiemann's coherent states) in the large $\eta$ limit \cite{Bianchi_2010}.

Despite the remarkable achievements in the standard (1+3)-dimensional LQG, the superstring theory in 10-dimensional spacetime, as another candidate of quantum gravity theory, shows significant advantages in unifying the gravity and the other three fundamental interactions. Hence, it is worth to explore LQG to the spacetime with dimension larger than four.  Pioneered by Bodendorfer, Thiemann and
Thurn \cite{Bodendorfer:Ha,Bodendorfer:La,Bodendorfer:Qu,Bodendorfer:SgI}, the issue of the loop quantum theory for general $(1+D)$-dimensional general relativity (GR) has been developed. The quantization bases on the connection formulation of $(1+D)$-dimensional GR in the form of the $SO(D+1)$ Yang-Mills theory, where the kinematic phase space consists of the spatial $SO(D+1)$ connection fields $A_{aIJ}$ and the vector fields $\pi^{bKL}$, endowed with the Poisson bracket $\{A_{aIJ}(x), \pi^{bKL}(y)\}=2\kappa\beta\delta_a^b\delta_{[I}^K\delta_{J]}^L\delta^{(D)}(x-y)$.  In this formulation, the dynamics is governed by a family of first class constraints which are the $SO(D+1)$ Gauss constraint, the $(1+D)$-dimensional ADM constraints and the additional constraint called the simplicity constraint. The first two types of constraints  are similar to that in the standard (1+3)-dimensional LQG. While the last one, i.e. the simplicity constraint, taking the form $S^{ab}_{IJKL}:=\pi^{a[IJ}\pi^{|b|KL]}$, generates the extra gauge symmetries in the $SO(D+1)$ Yang-Mills phase space. Once the Gauss and simplicity constraints are solved, this theory returns to the usual ADM formula of $(1+D)$-dimensional GR.  Similar to the standard LQG theory, the loop quantization of this $SO(D+1)$ formulation gives the spin-network states of the $SO(D+1)$ holonomies, carrying the quanta of the flux operators which represents the flux of $\pi^{bKL}$ over $(D-1)$ surfaces.

Even though the basic structure of $(1+D)$-dimensional LQG has been constructed \cite{long2020operators,Long:2020agv,Zhang:2015bxa}, the study on its coherent states is still very little \cite{Long:2020euh}. To construct a coherent state in $(1+D)$-dimensional LQG where gauge group $SO(D+1)$ is still compact, one could simply follow the Thiemann's procedure in (1+3)-dimensional case to construct a heat-kernel type coherent state.  However, because of the complicatedness of $SO(D+1)$, some specific calculations, like to study the peakedness and Ehrenfest properties, will be too difficult to proceed. This results in a huge obstacle for the study of the $SO(D+1)$ heat-kernels coherent states. However, as in the (1+3)-dimensional case, we can take advantage of the the coherent intertwiner in $(1+D)$-dimensional LQG to construct the superposition type coherent states.  Since superposition type coherent state in (1+3)-dimensional case can be related to the heat kernel one in large $\eta$ limit, we should expected that this relation can still be kept in the $(1+D)$-dimensional case. This paper mainly focus on the issue to construct the superposition type coherent state explicitly and prove the relation in the later case.

As shown in this paper below, to construct the superposition type coherent states  in $(1+D)$-dimensional LQG, the simplicity constraint should be imposed properly. More precisely, the elements in the constraint surface, solving the simplicity constraint, can be polar decomposed to adapt a geometrical parametrization. This parametrization gives a geometrical interpretation to each element and clarifies the gauge degrees of freedom associated to the simplicity constraint, which ensures the extension of superposition coherent state from $(1+3)$-dimensional case to the $(1+D)$-dimensional theory. Once the superposition coherent state is constructed, its properties are calculated and its relation to the  heat-kernel coherent state is studied. We find that the relation similar as that in (1+3)-dimensional case still exists.

This paper is organized as follows. In section 2, we will review the basic  structure of all dimensional LQG. We will first introduce the connection dynamics of GR in $(1+D)$-dimensional spacetime, upon which the holonomy-flux phase space will be introduced. Especially, we will emphasis the generalized twisted geometry parametrization of the holonomy-flux variables. Then, the spin network labelled by the coherent intertwiners will be pointed out as a special basis of the quantum Hilbert space. In section 3,  we will give the specific formulation of the superposition type coherent states in all dimensional LQG in both gauge variant and invariant case, and show that they are consistent with the Thiemann's coherent states for $SO(D+1)$ gauge theory in large $\eta$ limit. Finally in section 4 and 5, we will study the completeness and peakedness properties of the superposition type coherent states respectively. We will then conclude with the outlook for the possible next steps of the future research.
\section{Kinematic structure of all dimensional loop quantum gravity}
The phase space of the continuum connection dynamics of  (1+D)-dimensional GR consists of the canonical pairs of fields $(A_{aIJ}, \pi^{bKL})$ on a spatial D-dimensional manifold $\sigma$, where $A_{aIJ}$ is a $SO(D+1)$ valued connection and $\pi^{bKL}$ is the canonical conjugate variable.
The Poisson bracket between them is given by
\begin{equation}\label{Poisson1}
\{{A}_{aIJ}(x), \pi^{bKL}(y)\}=2\kappa\beta\delta_a^b\delta_{[I}^K\delta_{J]}^{L}\delta^{(D)}(x-y).
\end{equation}
This Hamiltonian system contains four constraints: the two kinematic constraints--Gauss constraint $\mathcal{G}^{IJ}\approx0$ and simplicity constraint $S^{ab[IJKL]}\approx0$, and the other two dynamics constraints---vector constraint $C_a\approx0$ and scalar constraint $C\approx0$. These constraints result in a first class constraint system.
 In the current paper, we will only need the explicit expression of the two  kinematic constraints which read
\begin{equation}
\text{Gauss \ constraint:}\quad\mathcal{G}^{IJ}\equiv \partial_a\pi^{aIJ}+2{A}_{aK}^{[I}\pi^{a|K|J]}\approx0
\end{equation}
and
\begin{equation}
\text{Simplicity \ constraint:}\quad S^{ab[IJKL]}=\pi^{a[IJ}\pi^{|b|KL]}\approx0
\end{equation}
As one expected, the Gauss constraint generates the $SO(D+1)$ gauge transformation, while the simplicity constraint restricts the degrees of freedom of $\pi^{aIJ}$ to that of a D-frame $E^{aI}$ which can be used to describe the spatial internal geometry. %The connection variables can be related to the geometric variables on the simplicity constraint surface.
More precisely, the solution of the simplicity constraint is given by $\pi^{aIJ}=2n^{[I}E^{|a|J]}$ with $E^{aI}$ being the densitized D-frame which gives the double densitized dual metric as $\tilde{\tilde{q}}^{ab}=E^{aI}E^b_I$. Moreover, $E^{aI}$ gives the D-bein $e_{aI}$ by $E^{aI}e_{bI}=\sqrt{q}\delta^a_b$ and then one can define the spin connection $\Gamma_{aIJ}$ by $\partial_{a}e_{b}^{I}-\Gamma_{ab}^ce_c^I+\Gamma_{a}^{IJ}e_{bJ}=0$ with  $\Gamma_{ab}^c$ being the Levi-Civita connection associated to $q_{ab}=e_{aI}e_b^I$. Based on these conventions, the densitized extrinsic curvature of the spatial manifold $\sigma$ can be given by
\begin{equation}
\tilde{{K}}_a^{\ b}\approx{ K}_{aIJ}\pi^{bIJ}\equiv \frac{1}{\beta}({A}_{aIJ}-\Gamma_{aIJ})\pi^{bIJ}
\end{equation}
which is a functional on the simplicity constraint surface. It is easy to see that ${K}_{aIJ}
:=\frac{1}{\beta}({A}_{aIJ}-\Gamma_{aIJ})$ can be decomposed as ${K}_{aIJ}=2n^{[I}{K}_a^{J]}+\bar{{K}}_{a}^{IJ}$, where $\bar{{K}}_{a}^{IJ}:=\bar{\eta}^I_K\bar{\eta}^J_L{K}_{a}^{KL}$ with $\bar{\eta}_I^J:=\delta_I^J-n_In^J$ and $\bar{{K}}_{a}^{IJ}n_I=0$. One can verify that the component $2n^{[I}{K}_a^{J]}$ is invariant and only $\bar{{K}}_{a}^{IJ}$ transforms under the gauge transformation generated by the simplicity constraint on the simplicity-constraint surface. Hence the simplicity constraint fixes both $\tilde{{K}}_a^{\ b}$ and $q_{ab}$ so that it exactly introduce extra gauge degrees of freedom.

The quantum geometry of loop quantum gravity is constructed by the spatially smeared variables---the conjugate momentum fluxes over surfaces and connection holonomies over paths--- for the conjugate pairs of elementary variables. The edges of the given graph naturally provide a set of paths to define holonomies, and the cell decomposition dual to the graph provides the set of (D-1)-faces specifying a fixed set of fluxes. In this setting, the holonomy over one of the edges is naturally conjugating to the flux over the face traversed by the edge, and the pairs associated with the given graph satisfy the smeared version of the algebra \eqref{Poisson1} and form a new phase space. More precisely, given a graph $\gamma$ embedded in the spatial manifold, we consider the new algebra given by
the discretized version of the connection $A_{aIJ}$ and its conjugate momentum $\pi^{aIJ}$. Namely, we consider the algebra consists of $(h_e, X_e)\in SO(D+1)\times so(D+1)$. Here
$h_e$ is given by $h_e=\mathcal{P}\exp\int_e \mathcal{A}$ with $\mathcal{P}$ denoting the path-ordered product, and $X_e$ is $X_e=\int_{e^\star}(h\pi h^{-1})^an_ad{}^{D-1}\!S$, where $e^\star$ is the dual (D-1)-dimensional face to the edge $e$ with the normal $n_a$ and infinitesimal coordinate area element $d{}^{D-1}\!S$, and $h$ is the parallel transport from one fixed vertex to the point of integration along a path adapted to the graph. Since $SO(D+1)\times so(D+1)\cong T^\ast SO(D+1)$, this new discrete phase space $\times_{e\in \gamma}(SO(D+1)\times so(D+1))_e$, called the phase space of $SO(D+1)$ loop quantum gravity on a fixed graph, is a direct product of $SO(D+1)$ cotangent bundles. Finally, the complete phase space of the theory is given by taking the union over the phase spaces of all possible graphs. Just like the $SU(2)$ case, the new variables $(h_e, X_e)$ of the phase space of $SO(D+1)$ loop quantum gravity can be seen as a discretized version of the continuum phase space. %离散variables的Poisson 括号
 Correspondingly, the discretized Gauss constraints read
 \begin{equation}
 G_v:=\sum_{b(e)=v}X_e-\sum_{t(e')=v}h_{e'}^{-1}X_{e'}h_{e'}\approx0
 \end{equation}
 and, the (discretized) simplicity constraints consisting of the edge-simplicity constraints $S^{IJKL}_e\approx0$ and vertex-simplicity constraints $S^{IJKL}_{v,e,e'}\approx0$, take the forms
\begin{equation}
\label{simpconstr}
S_e^{IJKL}\equiv X^{[IJ}_e X^{KL]}_e\approx0, \ \forall e\in \gamma,\quad S_{v,e,e'}^{IJKL}\equiv X^{[IJ}_e X^{KL]}_{e'}\approx0,\ \forall e,e'\in \gamma, s(e)=s(e')=v.
\end{equation}
The Poisson algebra between the variables $(h_e, X_e)$ is identical with the one given by the natural symplectic potential $\Theta_e=-\text{tr}(X_edh_eh^{-1}_e)$ on each $T^\ast SO(D+1)_e$. Based on this symplectic structure, one may evaluate the algebra amongst the discretized Gauss constraints, edge-simplicity constraints and vertex-simplicity constraints. It turns out that $G_v\approx0$ and $S_e\approx0$ form a first class constraint system, with the algebra
\begin{eqnarray}
\label{firstclassalgb}
\{S_e, S_e\}\propto S_e\,,\,\, \{S_e, S_v\}\propto S_e,\,\,\{G_v, G_v\}\propto G_v,\,\,\{G_v, S_e\}\propto S_e,\,\,\{G_v, S_v\}\propto S_v, \quad b(e)=v,
\end{eqnarray}
where the brackets within $G_v\approx0$ is just the $so(D+1)$ algebra, and the ones within $S_e\approx0$ weakly vanish. The algebra involving the vertex-simplicity constraint are the problematic ones, with the open anomalous brackets
\begin{eqnarray}
\label{anomalousalgb}
\{S_{v,e,e'},S_{v,e,e''}\}\propto \emph{anomaly term}
\end{eqnarray}
where the $ \emph{anomaly term}$ are not proportional to any of the existing constraints in the phase space.

The proper treatment of the anomalous simplicity constraints follows a reparametrization of the discrete phase space via the so-called generalized twisted geometries. The generalized twisted geometric parameters covers the degrees of freedom of the Regge geometries, thus the discrete phase space can get back to the geometrical dynamics phase space in some continuum limit \cite{PhysRevD.103.086016}. This reparametrization is given as follows. As mentioned above, the discrete phase space associated to a given graph $\gamma$ is $\times_{e\in \gamma}T^\ast SO(D+1)_e$. In this phase space, the edge simplicity constraint surface which we are interested in is given by
\begin{equation}
\times_{e\in \gamma}T_{\text{s}}^\ast SO(D+1)_e:=\{(h_e,X_e)\in \times_{e\in \gamma}T^\ast SO(D+1)_e|X_{e}^{[IJ}X_{e}^{KL]}=0\}.
\end{equation}
Moreover, the angle-bivector parametrization of the edge-simplicity constraint surface $T_{\text{s}}^\ast SO(D+1)$ for a single edge is based on the generalized twisted-geometry variables $(V,\tilde{V},\xi^o, N^o,\bar{\xi}^\mu)\in P:=Q_{D-1}\times Q_{D-1}\times T^*S^1\times SO(D-1)$, where the bi-vectors $V$ or $\tilde{V}$ constitutes the space $Q_{D-1}:=SO(D+1)/(SO(2)\times SO(D-1))$ with $SO(2)\times SO(D-1)$ being the maximum subgroup in $SO(D+1)$ preserving the bivector $\tau_o:=2\delta_1^{[I}\delta_2^{J]}$,  $N^o\in \mathbb{R}$, $\xi^o\in [0,2\pi)$ with $e^{\xi^o\tau_o}\in SO(2)$ and $e^{\bar{\xi}^\mu\bar{\tau}_\mu}:=\bar{u}\in SO(D-1)$, $\mu\in\{1,...,\frac{(D-1)(D-2)}{2}\}$ with $\bar{\tau}_\mu$ being a basis of $so(D-1)$. To capture the intrinsic curvature, we have specified one pair of the $SO(D+1)$ valued Hopf sections $u(V)$ and $\tilde{u}( \tilde{V})$ satisfying $V=u(V)\tau_ou(V)^{-1}$ and $\tilde{V}=-\tilde{u}(\tilde{V})\tau_o\tilde{u}(\tilde{V})^{-1}$. With the specified $u(V)$ and $\tilde{u}( \tilde{V})$, the parametrization associated with each edge is given by the map
\begin{eqnarray}\label{para}
P\ni(V,\tilde{V},\xi^o,N^o,\bar{\xi}^\mu)\mapsto(h, X)\in T_{\text{s}}^\ast SO(D+1):&& X=N^o\,V=N^o\,u(V)\tau_ou(V)^{-1}\\\nonumber
&&h=u(V)\,e^{\bar{\xi}^\mu\bar{\tau}_\mu}e^{\xi^o\tau_o}\,\tilde{u}(\tilde{V})^{-1}.
\end{eqnarray}
 This map is a two-to-one double covering of the image that takes the bi-vector form $X=N^ou\tau_ou^{-1}$ solving the edge-simplicity constraint $X^{[IJ}X^{KL]}=0$. More explicitly, under this map from $P$ to $T_{s}^\ast \!SO(D+1)$, the two points $(V,\tilde{V},\xi^o, N^o,\bar{\xi}^\mu)$ and $(-V,-\tilde{V},-\xi^o,-N^o,\dot{\xi}^\mu)$ related by $e^{\dot{\xi}^\mu\bar{\tau}_{\mu}}=e^{-2\pi\tau_{13}}e^{\bar{\xi}^\mu\bar{\tau}_{\mu}}e^{2\pi\tau_{13}}$ and $\tau_{13}=\delta_1^{[I}\delta_3^{J]}$ are mapped to the same point $(h, X)\in T_{s}^\ast \!SO(D+1) $. A bijection map can thus be established in the region $|X|\neq0$ by selecting either branch among the two signs, see more details in \cite{PhysRevD.103.086016}. Also, the Poisson structures given by the new parameters can be much simplified, i.e., we have the non-vanishing Poisson bracket
 \begin{equation}\label{xiN}
 \{\xi^o, N^o\}\propto 1
 \end{equation}
 with $\xi^o$ and $N^o$ capture the degrees of freedom in extrinsic and intrinsic geometry respectively.
     This parametrization for the phase space based on one edge $e$ can be extended to the whole graph $\gamma$ so that the Gauss constraint and vertex simplicity constraint can be imposed at the vertices. A remarkable result based on this parametrization is that, on the constraint surface of both edge simplicity and anomalous vertex simplicity constraints in discrete phase space, the gauge transformation induced by the edge simplicity constraint is exactly identical with the gauge transformation induced by the non-anomalous simplicity constraint in continuum connection phase space in continuum limit. This relation can be illustrated as
  \begin{equation}
\bar{\xi}^\mu_e\stackrel{\textrm{continuum limit}}{---\rightarrow} \bar{K}_{aIJ}
  \end{equation}
  where $\bar{\xi}^\mu_e$ takes the pure gauge degrees of freedom with respect to simplicity constraint in holonomy $h_e$.
  Hence, solving the kinematic constraints follows  the symplectic reduction with respect to edge simplicity constraint and Gauss constraint in the original discrete phase space, with the vertex simplicity constraint being imposed weakly. It has been shown that,  the resulted constrained discrete space gives the generalized twisted geometry space covering the degrees of freedom of internal and external Regge geometry on the D-dimensional spatial manifold $\sigma$.
\subsection{Spin network basis and simple coherent intertwiner in all dimensional loop quantum gravity}

The Hilbert space of all dimensional LQG is given by the completion of the space of cylindrical functions on quantum configuration space, which can be decomposed into the sectors --- the Hilbert spaces constructed on graphs. For a given graph $\gamma$ with $|E(\gamma)|$ edges, the related Hilbert space is given by $\mathcal{H}_\gamma=L^2(SO(D+1)^{|E(\gamma)|}, d\mu^{|E(\gamma)|})$. This Hilbert space associates to the classical phase space $\times_{e\in\gamma}T^\ast SO(D+1)_e$ aforementioned. A basis of this space is given by the spin-network functions which are labelled by (1) an $SO(D+1)$ representation $\Lambda$ assigned to each edge; and (2) an intertwiner $i_v$ assigned to each vertex $v$. Each basis state $\Psi_{\gamma,{\vec{\Lambda}}, \vec{i}}(\vec{h}_e)$, as a wave function on $\times_{e\in\gamma}SO(D+1)_e$, is then given by
\begin{eqnarray}
\Psi_{\gamma,{\vec{\Lambda}}, \vec{i}}(\vec{h}_e)\equiv \bigotimes_{v\in\gamma}{i_v}\,\, \rhd\,\, \bigotimes_{e\in\gamma} \pi_{\Lambda_e}(h_{e}),
\end{eqnarray}
where $\vec{h}_e:=(...,h_e,...), \vec{\Lambda}:=(...,\lambda_e,...), e\in\gamma$, $\vec{i}:=(...,i_v,...), v\in\gamma$ , $\pi_{\Lambda_e}(h_{e})$ denotes the holonomy $h_e$ associated to edge $e$ in the representation labelled by $\Lambda_e$, and $\rhd$ denotes the contraction of the intertwiners with the representation matrixes of holonomies. The wave function is then simply the product of the specified components of the holonomy matrices, selected by the intertwiners at the vertices. The kinematic physical states can be obtained by solving the kinematic constraints. The strong imposition of edge simplicity constraint and Gauss constraint gives the edge-simple and gauge invariant spin network states, whose edges are labelled by the simple representations of $SO(D+1)$ and vertices are labelled by the gauge invariant intertwiners. The anomalous vertex simplicity constraints are imposed weakly and the corresponding weak solutions is given by the spin network states labelled by the simple coherent intertwiners \cite{long2019coherent}. A typical spin network state $\Psi_{\gamma,\vec{N},\vec{\mathcal{I}}_{\text{s.c.}}}(h_e(A))$ labelled by the gauge invariant simple coherent intertwiners $\mathcal{I}_{\text{s.c.}}$ is given by
\begin{equation}
\Psi_{\gamma,\vec{N},\vec{\mathcal{I}}_{\text{s.c.}}}(\vec{h}_e(A))=\text{tr}(\otimes_{e\in\gamma} \pi_{N_e}(h_e(A))\otimes_{v\in\gamma}\mathcal{I}_v^{\text{s.c.}})
\end{equation}
where $\pi_{N_e}(h_e(A))$ represents the representation matrix of $h_e(A)\in SO(D+1)$ in the simple representation space labelled by an non-negative integer $N_e$, and $\mathcal{I}_v^{\text{s.c.}}$ is the so-called simple coherent intertwiner.

 Before giving the explicit expression of simple coherent intertwiners,  let us first introduce some basic concepts of the homogeneous harmonic functions on the $D$-sphere.
 The homogeneous harmonic functions with degree $N$ on the $D$-sphere provide an irreducible representation space of $SO(D+1)$, denoted by $\mathfrak{H}_{D+1}^{N}$ with dimensionality $\dim(\mathfrak{H}_{D+1}^{N})=\dim(\pi_N)=\frac{(D+N-2)!(2N+D-1)}{(D-1)!N!}$. Introduce a subgroup series $SO(D+1)\supset SO(D)\supset SO(D-1)\supset ... \supset SO(2)_{\delta_1^{[I}\delta_2^{J]}}$ where $SO(2)_{\delta_1^{[I}\delta_2^{J]}}$ is the one-parameter subgroup of SO$(D+1)$ generated by $\tau_o:=2\delta_1^{[I}\delta_2^{J]}$. An orthogonal basis of the space $\mathfrak{H}_{D+1}^{N}$ in Dirac bra-ket notation can be denoted by $\ket{N,\mathbf{M}}$ where $\mathbf{M}:=M_1,M_2,...,M_{D-1}$ with $N\geq M_1 \geq M_2\geq...\geq|M_{D-1}|$, and $N,M_1,... M_{D-2}\in\mathbb{N}$, $M_{D-1}\in \mathbb{Z}$. The labels $N, \mathbf{M}$ can be interpreted through that $\ket{N,\mathbf{M}}$ belongs to the series of space $\mathfrak{H}_{2}^{M_{D-1}}\subset \mathfrak{H}_{3}^{M_{D-2}}\subset...\subset\mathfrak{H}_{D}^{M_{1}}\subset\mathfrak{H}_{D+1}^{N}$ which are the irreducible representation spaces labeled by $M_{D-1},...,M_2,M_1, N$ of the series of groups $SO(2)_{\delta_1^{[I}\delta_2^{J]}} \subset SO(3)\subset ... \subset SO(D)\subset SO(D+1)$ respectively \cite{vilenkin2013representation}. Based on this convention, the corresponding inner product reads
\begin{equation}
\langle N,\mathbf{M}|N,\mathbf{M}'\rangle:=\delta_{\mathbf{M},\mathbf{M}'}
\end{equation}
with $\delta_{\mathbf{M},\mathbf{M}'}=1$ if $\mathbf{M}=\mathbf{M}'$ and zero otherwise. The state in $\mathfrak{H}_{D+1}^{N}$ labelled by the highest weight vector is $|N_e,\mathbf{N}_{e}\rangle$ with $\mathbf{N}_{e}=\mathbf{M}|_{M_1=...=M_{D-1}=N_{e}}$, and the corresponding Perelomov type coherent states in this space is defined by $|N,V\rangle:=g|N,\mathbf{N}\rangle$ with $g\in SO(D+1)$ and $V=g\tau^og^{-1}$ \cite{Long:2020euh}.%%%V g N 的关系.

Now let us consider the coherent intertwiner at a vertex $v$. Without loss of generality, we re-orient the edges linked to $v$ to be outgoing at $v$.  The gauge fixed coherent intertwiners, as elements of the space $\mathcal{H}^{\vec{N}_e}_v:=\otimes_{ b(e)=v}\overline{\mathfrak{H}}^{N_{e},D+1}$, are defined as  $\check{\mathcal{I}}_v^{\text{c.}}(\vec{N},\vec{V}):=\otimes_{e: b(e)=v}\langle N_e,V_e|$, where $\bar{\mathfrak{H}}^{N_{e},D+1}$ is the dual space of homogeneous harmonic functions with degree $N_e$ on the D-sphere and $|N_{e},V_{e}\rangle:=u(V_{e})|N_{e},\mathbf{N}_{e}\rangle$ with $u(V_{e})$ being specific $SO(D+1)$ valued function of $V_e$ satisfying $V_e=u(V_{e})\tau_ou(V_{e})^{-1}$. Specifically, the gauge invariant coherent intertwiners $\mathcal{I}_v^{\text{c.}}$ are given by the group averaging of $\otimes_{e: b(e)=v}\langle N_e,V_e|$, namely $\mathcal{I}_v^{\text{c.}}(\vec{N},\vec{V}):=\int_{SO(D+1)}dg\otimes_{e: b(e)=v}\langle N_e,V_e|g$. Specifically, the so-called simple coherent intertwiners $\check{\mathcal{I}}_v^{\text{s.c.}}$ (or $\mathcal{I}_v^{\text{s.c.}}$ in gauge invariant case) is defined by requiring $V_e^{[IJ}V_{e'}^{KL]}=0$ with $b(e)=b(e')=v$ in their definition. It has been shown that, the simple coherent intertwiners weakly solve the vertex simplicity constraint, and they capture correct spatial geometric degrees of freedom in large $N$ limit, which means that the weak imposition servers as a proper treatment for the anomalous vertex simplicity constraint \cite{long2019coherent}.

\section{The super position type coherent states in all dimensional LQG}
%Though the heat-kernel coherent state is widely used in the study of standard (1+3)-dimensional LQG, another type of coherent states is firstly introduced by Rovelli and further studied in [][], which is referred to as the super position type coherent states.
Thanks to  the coherent intertwiner constructed in all dimensional LQG, this super position type coherent states can be generalized to all dimensional cases directly. For a given graph $\gamma$, we define the gauge variant super position type coherent states in all dimensional LQG as
\begin{equation}\label{spcs}
\tilde{\Psi}_{\gamma,\vec{\mathbb{H}}^o_{e}}(\vec{h}_{e}):= \sum_{N_{e}}\left(\prod_{e}\dim(\pi_{N_{e}}) \exp(-\frac{(N_{e}-N_{e}^o)^2}{2\sigma^0_{e}})e^{-\mathbf{i}\xi^o_{e}N_{e}}\right) \cdot\Psi_{\gamma,\vec{N}_{e},\vec{\check{\mathcal{I}}}_{v}(\vec{V}_{e_v})}(\vec{h}_{e})
\end{equation}
and  the gauge invariant one, as
\begin{equation}\label{gispcs} \tilde{\underline{\Psi}}_{\gamma,\vec{\mathbb{H}}^o_{e}}(\vec{h}_{e}):=\sum_{N_{e}}\left(\prod_{e}\dim(\pi_{N_{e}}) \exp(-\frac{(N_{e}-N_{e}^o)^2}{2\sigma^0_{e}})e^{-\mathbf{i}\xi^o_{e}N_{e}}\right) \cdot\underline{\Psi}_{\gamma,\vec{N}_{e},\vec{{\mathcal{I}}}_{v}(\vec{V}_{e_v})}(\vec{h}_{e}).
\end{equation}
where $\vec{\mathbb{H}}:=\{\mathbb{H}_e^o\}_{e\in E(\gamma)}$ with $\mathbb{H}_e^o=(V_e, \tilde{V}_e, N^o_e, \xi_e^o)$. Note that such kind of coherent states are defined to be independent with the pure gauge component $\bar{\xi}^\mu_e$.
%解释\mathbb{H}和H的关系，抹掉规范自由度。
%It is noted that $\mathbb{H}_e^o$, comparing to $H^o$ defined in \eqref{spcs}, is independent of $\bar{g} \bar{\tilde{g}}^{-1}=e^{\bar{\xi}^\mu\bar{\tau}_\mu}$. The reason why the labels of the super position type coherent states are defined to be $\mathbb{H}^0_e$ rather than $H^0_e$ will be discussed in details below.
In the following parts of this section, we will show that the super position type coherent state is identical with the heat kernel one in all dimensional LQG in large $\eta$ limit.
\subsection{Heat kernel coherent states of $SO(D+1)$}
Now let us give some details of the heat kernel coherent states in $\mathcal{H}_\gamma$, which are labeled by points in the phase space $\times_{e\in\gamma}T^\ast SO(D+1)_e$. These coherent states are tensor product of coherent states associated to edges $e\in \gamma$, and the coherent states associated to each edge are the heat kernel coherent states of $SO(D+1)$. The  heat kernel coherent states of $SO(D+1)$ is given by
\begin{equation}\label{eq:kt}
  K_t(h,H)=\sum_{\Lambda}\dim(\pi_{\Lambda})e^{t\Delta}\chi^{\pi_{\Lambda}}(hH^{-1}),
\end{equation}
where $H\in SO(D+1)_{\mathbb{C}}\cong T^\ast SO(D+1)$, $\pi_{\Lambda}$ is the representation of $SO(D+1)$ labelled by its highest weight vector $\Lambda$, $-\Delta$ is the Casimir operator of $SO(D+1)$, and $\chi^{\pi_{N}}(hH)$ represents the trace of $hH$ in the representation $\pi_{N}$. In all dimensional LQG, the simplicity constraint restricts the representations of holonomies to be the simple ones. Thus, we vanish those terms in \eqref{eq:kt} corresponding to the non-simple representations. We also note that the super position type coherent states are labelled by the points on the edge simplicity constraint surface, in order to discuss their relations,  we also restrict the label $H$ of heat kernel coherent states to be $H^o$ which takes values in the edge simple constraint surface $SO(D+1)^{\text{s.}}_{\mathbb{C}}\cong T^\ast_{\text{s.}}SO(D+1)$. This procedure gives the simple heat kernel coherent states of $SO(D+1)$ as,
 \begin{equation}
  K_t(h,H^o)=\sum_{N}\dim(\pi_{N})e^{-N(N+D-1)t}\chi^{\pi_{N}}(h{H_e^o}^{-1}),
\end{equation}
where $\pi_{N}$ is the simple representation of $SO(D+1)$ labelled by non-negative integer $N$.
Based on the polar decomposition of $SO(D+1)_{\mathbb{C}}$, each element $H^o\in SO(D+1)^{\text{s.}}_{\mathbb{C}}$ can be written in terms of a positive real number $\eta$ and
two independent $SO(D+1)$ group elements $g$ and $\tilde{g}$ as
\begin{equation}\label{pd}
H^o=g\exp{\left(\mathbf{i}\eta\tau_o\right)}\tilde{g}^{-1}.
\end{equation}
Given an fixed choice of Hopf section $u(V)$ in $SO(D+1)$, an element $g\in SO(D+1)$ can be uniquely decomposed as
\begin{equation}
g=u(V)e^{\phi\tau_o}\bar{g}
\end{equation}
with an angle $\phi$, an element $\bar{g}$ of $SO(D-1)$ preserving  $\tau_o$ and an unit bi-vector $V\in Q_{D-1}$ satisfying $V=u(V)\tau_ou^{-1}(V)$.
Taking advantage of this expression, we finally decompose $H^o$ as
\begin{equation}
H^o=u(V)e^{\phi\tau_o}\bar{g}\exp{\left(\mathbf{i}\eta\tau_o\right)} e^{-\tilde{\phi}\tau_o}\bar{\tilde{g}}^{-1}\tilde{u}^{-1}(\tilde{V})
=u(V)\bar{g} \bar{\tilde{g}}^{-1}\exp{\left(z\tau_o\right)} \tilde{u}^{-1}(\tilde{V}),
\end{equation}
with $z=(\phi-\tilde{\phi})+\mathbf{i}\eta=:\xi^o+\mathbf{i}\eta$, $\bar{g}, \bar{\tilde{g}}\in SO(D-1)$, $u(V), \tilde{u}(\tilde{V}) \in Q_{D-1}$. This decomposition repeats the parametrization of $T^\ast_{\text{s}} SO(D+1)$ by $(N^o, V,\tilde{V},\xi^o,\bar{\xi}^\mu)$ introduced in section 2, where $\eta$ is proportion to $N^o$ and $\bar{g} \bar{\tilde{g}}^{-1}=e^{\bar{\xi}^\mu\bar{\tau}_\mu}$.
\subsection{Relation between two kinds of coherent states}
\subsubsection{Gauge variant formulation}
By definition, the Heat kernel coherent state on a graph $\gamma$ reads
\begin{equation}\label{heatcoherent}
\Psi_{\gamma,\vec{H}^o_{e}}(\vec{h}_{e})=\prod_{e\in\gamma}K_e^{t_e}(h_e,H_e^o) =\prod_{e\in\gamma}\sum_{N_e}\dim(\pi_{N_e})e^{-N_e(N_e+D-1)t_e}\chi^{\pi_{N_e}}(h_e{H_e^o}^{-1}).
\end{equation}
%This state can be expanded on the spin-network basis $\Psi_{\gamma,\vec{N}_{e},\vec{i}_v}(\vec{h}_{e})$ as
%\begin{equation}
%\Psi_{\gamma,\vec{H}^o_{e}}(\vec{h}_{e})=\sum_{\vec{N}_{e}}\sum_{\vec{i}_v} f_{\vec{N}_{e},\vec{i}_v,\vec{H}^o_{e}}\Psi_{\gamma,\vec{N}_{e},\vec{i}_v}(\vec{h}_{e}),
%\end{equation}
%with the heat kernel components $f_{\vec{N}_{e},\vec{i}_v,\vec{H}^o_{e}}$ given by
%\begin{equation}
%f_{\vec{N}_{e},\vec{i}_v,\vec{H}^o_{e}}=\text{tr}\left(\prod_{e}\dim(\pi_{N_{e}}) e^{-N_{e}(N_{e}+D-1)t_{e}}D^{(\pi_{N_{e}})}(H^o_{e}) \cdot\prod_{v}i_v\right),
%\end{equation}
%where $D^{(\pi_{N_{e}})}(H^o_{e})$ is the representation matrix of $H^o_{e}$ in the representation of $SO(D+1)$ labelled by $N_e$.
For the curious cases with $\eta_{e}\gg1$, the matrix $\exp{\left(-z_{e}\tau_o\right)}$ in the decomposition of ${H_e^o}^{-1}$ under a proper basis can be simplified as
\begin{equation}\label{etagg1}
\langle N_{e},\mathbf{M}|\exp{\left(-z_{e}\tau_o\right)}|N_{e},\mathbf{M}'\rangle =\delta^{\mathbf{M}}_{ \ \mathbf{M}'}e^{-\mathbf{i}z_{e}M_{D-1}} =\delta^{\mathbf{M}}_{\ \mathbf{M}'}\exp{\left(\eta_{e}N_{e}\right)} \left(\delta_{\mathbf{M}, \mathbf{N}_{e}}e^{-\mathbf{i}\xi^o_{e}N_{e}}+\mathcal{O}(e^{-\eta_{e}})\right),
\end{equation}
where $\mathbf{N}_{e}=\mathbf{M}|_{M_1=...=M_{D-1}=N_{e}}$. Thus, by introducing the projector $P_{\text{h.}}=|N_{e},\mathbf{N}_{e}\rangle\langle N_{e},\mathbf{N}_{e}|$ which project each vector to be parallel to $|N_{e},\mathbf{N}_{e}\rangle$, we have
\begin{equation}\label{etagg2}
\sum_{\mathbf{M},\mathbf{M}'}|N_{e},\mathbf{M}\rangle\langle N_{e},\mathbf{M}|\exp{\left(-z_{e}\tau_o\right)}|N_{e},\mathbf{M}'\rangle\langle N_{e},\mathbf{M}'| \approx e^{\eta_{e}N_{e}} e^{{-\mathbf{i}}\xi^o_{e}N_{e}}P_{\text{h.}}.
\end{equation}
 Let us insert Eq.\eqref{etagg2} into ${H_e^o}^{-1}$ and notice that $\bar{g}_{e} \bar{\tilde{g}}^{-1}_{e}|N_{e},\mathbf{N}_{e}\rangle=|N_{e},\mathbf{N}_{e}\rangle$, $u(V_{e})|N_{e},\mathbf{N}_{e}\rangle=|N_{e},V_{e}\rangle$ and $\tilde{u}(\tilde{V}_{e})|N_{e},\mathbf{N}_{e}\rangle=|N_{e},-\tilde{V}_{e}\rangle$ by their definition, we can get
 \begin{equation}\label{DH}
D^{(\pi_{N_{e}})}({H_e^o}^{-1})\approx e^{\eta_{e}N_{e}} e^{{-\mathbf{i}}\xi^o_{e}N_{e}}|N_{e},-\tilde{V}_{e}\rangle\langle N_{e},V_{e}|,
 \end{equation}
 where $D^{(\pi_{N_{e}})}(H^o_{e})$ is the representation matrix of $H^o_{e}$ in the representation of $SO(D+1)$ labelled by $N_e$.
% Let us introduce the dual gauge fixed coherent intertwiner $\check{\mathcal{I}}_{v}(\vec{N}_{e_v},\vec{V}_{e_v}):=
%\otimes_{t(e) =v}|N_{e},V_{e}\rangle
%\otimes_{t(e) =v}\langle N_{e},V_{e}|$ in the dual intertwiner space $\bar{\mathcal{H}}^{\vec{N}_e}_v:=%\otimes_{ t(e)=v}\mathfrak{H}^{N_{e},D+1}
%\otimes_{ t(e)=v}\bar{\mathfrak{H}}^{N_{e},D+1}$ at vertex $v$ with all of the edges linked to $v$ being oriented ingoing at $v$, where $e_v\in\{e| t(e)=v\}$ and $\bar{\mathfrak{H}}^{N_{e},D+1}$ is the dual space of $\mathfrak{H}^{N_{e},D+1}$. The dual coherent intertwiner $\check{\mathcal{I}}_{v}(\vec{N}_{e_v},\vec{V}_{e_v})$ have components on an orthonormal basis $\{i_v\}$ of the intertwiner space at $v$, which can be shown as
%$\check{ \mathcal{I}}_{v}(\vec{N}_{e_v},\vec{V}_{e_v}) =\sum_{i_v}\check{\mathcal{I}}_{i_v}(\vec{N}_{e_v},\vec{V}_{e_v})i_v$, with the components being given by
%\begin{equation}
%\check{\mathcal{I}}_{i_v}(\vec{N}_{e_v},\vec{V}_{e_v}) =\text{tr}(i_v\cdot\check{\mathcal{I}}_{v}(\vec{N}_{e_v},\vec{V}_{e_v})).
%\end{equation}
Then, let us insert the expression \eqref{DH} of $D^{(\pi_{N_{e}})}({H_e^o}^{-1})$ into Eq.\eqref{heatcoherent} and notice that
\begin{equation}\label{neta}
-N_e(N_e+D-1)t_e+N_e\eta_e=-(N_e-\frac{\eta_e-t_e(D-1)}{2t_e})^2t_e+\frac{(\eta_e-t_e(D-1))^2}{4t_e},
\end{equation}
we find that the large $\eta_{e}$ limit of the heat kernel coherent spin-networks \eqref{heatcoherent}, which is given by the super-position type coherent state
\begin{equation}
\tilde{\Psi}_{\gamma,\vec{\mathbb{H}}^o_{e}}(\vec{h}_{e}):= \sum_{N_{e}}\left(\prod_{e}\dim(\pi_{N_{e}}) \exp(-\frac{(N_{e}-N_{e}^o)^2}{2\sigma^0_{e}})e^{-\mathbf{i}\xi^o_{e}N_{e}}\right) \cdot\Psi_{\gamma,\vec{N}_{e},\vec{\check{\mathcal{I}}}_{v}(\vec{V}_{e_v})}(\vec{h}_{e})
\end{equation}
up to a normalization factor, where we defined
\begin{equation}\label{etat}
\sigma^0_{e}\equiv\frac{1}{2 t_{e}},\quad N^o_{e}\equiv\frac{\eta_e-t_e(D-1)}{2t_e}
\end{equation}
and introduced the spin-network states $\Psi_{\gamma,\vec{N}_{e},\vec{\mathcal{I}}_{v}(\vec{V}_{e_v})}(\vec{h}_{e}) $ whose vertices being labelled by gauge fixed coherent intertwiners $\check{\mathcal{I}}_{v}(\vec{N}_{e_v},\vec{V}_{e_v})$ defined by
\begin{equation}
 \check{\mathcal{I}}_{v}(\vec{N}_{e_v},\vec{V}_{e_v}):=
\otimes_{t(e) =v}|N_{e},-\tilde{V}_{e}\rangle
\otimes_{b(e) =v}\langle N_{e},V_{e}|
\end{equation}
 Note that the original heat kernel coherent states  \eqref{heatcoherent} are labeled by the $H^o_{e}=u(V_{e})\bar{g}_{e} \bar{\tilde{g}}^{-1}_{e}\exp{\left(z_{e}\tau_o\right)} \tilde{u}^{-1}(\tilde{V}_{e})$ which is parametrized by $(V_e, \tilde{V}_e, \bar{g}_{e} \bar{\tilde{g}}^{-1}_{e}, z_{e}=\xi^o_{e}+\mathbf{i}\eta_{e})$, while the super-position type coherent state \eqref{spcs}
is labelled by the parameters $\mathbb{H}_e^o=(V_e, \tilde{V}_e, N^o_e, \xi_e^o)$ only. As we have mentioned, the degrees of freedom of $\bar{g}_{e} \bar{\tilde{g}}^{-1}_{e}$ among these parameters is exactly the gauge degrees of freedom with respect to the simplicity constraint, hence the large $\eta$ limit removes the dependence of heat kernel coherent states on these gauge degrees of freedom correctly.

\subsubsection{Gauge invariant formulation}
The Gaussian constraint in all dimensional LQG requires that the physical quantum states should be gauge invariant with respect to the $SO(D+1)$ rotation induced by Gaussian constraint. Hence, it is worth to consider the same expansion of the gauge invariant heat-kernel coherent states $\underline{\Psi}_{\gamma,\vec{H}^o_{e}}(\vec{h}_{e})$ on spin-network basis. Such states can be given by projecting the heat-kernel coherent states ${\Psi}_{\gamma,\vec{H}^o_{e}}(\vec{h}_{e})$ onto the gauge invariant space by group averaging, which reads,
\begin{eqnarray}\label{ginvhk}
\underline{\Psi}_{\gamma,\vec{H}^o_{e}}(\vec{h}_{e})&=&\int\prod_{v\in\gamma}dg_v\prod_{e\in\gamma}K_e^t(h_e,g_{b(e)}H_e^og^{-1}_{t(e)}) \\\nonumber
&=&\int\prod_{v\in\gamma}dg_v\prod_{e\in\gamma}\sum_{N_e}\dim(\pi_{N_e})e^{-N_e(N_e+D-1)t}\chi^{\pi_{N_e}}(h_e(g_{b(e)}H_e^og^{-1}_{t(e)})^{-1}).
\end{eqnarray}
Similar to the gauge variant case, by taking the same asymptotics for $\eta_{e}\gg1$,  we can substitute Eq.\eqref{DH} into the gauge invariant heat kernel coherent state \eqref{ginvhk} and find the following asymptotics
\begin{equation} \tilde{\underline{\Psi}}_{\gamma,\vec{\mathbb{H}}^o_{e}}(\vec{h}_{e}):=\sum_{N_{e}}\left(\prod_{e}\dim(\pi_{N_{e}}) \exp(-\frac{(N_{e}-N_{e}^o)^2}{2\sigma^0_{e}})e^{-\mathbf{i}\xi^o_{e}N_{e}}\right) \cdot\Psi_{\gamma,\vec{N}_{e},\vec{{\mathcal{I}}}_{v}(\vec{V}_{e_v})}(\vec{h}_{e})
\end{equation}
up to a normalization factor, which is exactly the gauge invariant coherent spin-networks of superposition type.

\section{Resolution of the identity}
It is easy to conclude that the $SO(D+1)$ heat kernel coherent states provides a holomorphic representation for all dimensional LQG by the results in \cite{Thomas2001Gauge}\cite{1994The}. However, with the same reason that the $SO(D+1)$ Heat kernel coherent states are too complicated for explicit computations, this result seems to make no too much sense for further studies. Hence, we expect that the more practical one --- the superposition type coherent states in this paper could make a resolution of the identity in the Hilbert space of all dimensional LQG. Notice that the superposition type coherent state have a complete Gaussian distributional factor only in the large $N^o$ limit. We will show that the superposition type coherent states with small and negative $N^o$ are also necessary to give a complete basis of the Hilbert space of all dimensional LQG, though they deviate from the Gaussian distribution so that they are not well coherent states comparing with the ones with large $N^o$.

Let us recall that the superposition type coherent states are labelled by $\mathbb{H}^o=(V, \tilde{V}, N^o, \xi^o)$, while the heat kernel coherent states are labelled by $H^o$ which can be parametrized by $(V, \tilde{V}, N^o, \xi^o)$ and an additional $SO(D-1)$ element through a two to one map.  More explicitly, the elements $(V, \tilde{V}, N^o, \xi^o)$ take values in the constitute space $Q_{D-1}\times Q_{D-1}\times T^\ast S^1$, with $Q_{D-1}\times Q_{D-1}\times T^\ast S^1\times SO(D-1)$ being a double cover of the discrete phase space (on an edge) $T^\ast_{\text{s}}SO(D+1)\ni H^o$ at $N^o\neq0$, and the $SO(D-1)$ components corresponding to the gauge degrees of freedom with respected to the simplicity constraint in $T^\ast_{\text{s}}SO(D+1)$. The double covering ensure that the range of $\mathbb{H}^o=(V, \tilde{V}, N^o, \xi^o)$
 is given by $Q_{D-1}\times Q_{D-1}\times T^\ast S^1$ so that $N^o$ takes value in $\mathbb{R}$. As we will see, the range of $N^o$ is a key point to give the resolution of the identity in the cylindrical space $\mathcal{H}_\gamma$ on graph $\gamma$ based on the superposition type coherent states.

 %The coherent states $\tilde{\Psi}_{\gamma,\vec{\mathbb{H}}^o}$ are labelled by the element $\vec{\mathbb{H}}^o$ of the space $\times_{e\in\gamma}(Q_{D-1}\times Q_{D-1}\times T^\ast S^1)_e$ which covers the kinematic physical degrees of freedom in the discrete phase space $\times_{e\in\gamma}(T^\ast_{\text{s}}SO(D+1))_e$. Hence, it is naturally to require that $\tilde{\Psi}_{\gamma,\vec{\mathbb{H}}^o}$ serves a resolution of the identity in the kinematic Hilbert space $\mathcal{H}_\gamma$ on graph $\gamma$. We will discuss it in both gauge (with respected to Gaussian constraint) fixing and gauge invariant case respectively.

We firstly consider the gauge fixed case.  Before turn to give the resolution of the identity in the kinematic Hilbert space $\mathcal{H}_\gamma$, let us introduce the identity of the intertwiner space by the coherent intertwiners.  Note the identity $\mathbb{I}_{\overline{\mathfrak{H}}^{N,D+1}}=\int_{Q_{D-1}}dV_N|N, V\rangle\langle N ,V|$ in the space $\overline{\mathfrak{H}}^{D+1,N}$ with $\int_{Q_{D-1}}dV=1$ and $dV_N=\dim(\pi_{N})\cdot dV$, we have that
\begin{equation}\label{id222}
\mathbb{I}_{{\mathcal{H}}^{\vec{N}_v}_v}= \int_{\mathcal{P}_v}d\vec{V}_{\vec{N}_{e_v}}|\check{\mathcal{I}}_{v}(\vec{N}_{e_v},\vec{V}_{e_v}) \rangle\langle \check{\mathcal{I}}_{v}(\vec{N}_{e_v},\vec{V}_{e_v})|
\end{equation}
is the identity in the intertwiner space $ \mathcal{H}^{\vec{N}_e}_v:=\otimes_{ b(e)=v}\overline{\mathfrak{H}}^{N_{e},D+1}$ at $v$, where the edges linked to $v$ are re-oriented to be outgoing at $v$ without loss of generality, the measure $d\vec{V}_{\vec{N}_{e_v}}:=\prod_{e:b(e)=v}dV_{\!N_{e}}$ is compatible with the natural symplectic structure of the phase space $\mathcal{P}_v:=\times_{e:b(e)=v}Q_{D-1}^{e}$ \cite{PhysRevD.103.086016}.
 %Correspondingly, we have
%\begin{eqnarray}
%\mathbb{I}_{\oplus_{\vec{N}_{e_v}}\mathcal{H}^v_{\vec{N}_{e_v}}}&=&\oplus _{\vec{N}_{e_v}}\int_{\times_{e_v}Q_{D-1}^{e_v}}d\vec{V}_{\!\vec{N}_{e_v}}|\mathcal{I}_v(\vec{V}_{e_v})\rangle\langle \mathcal{I}_v(\vec{V}_{e_v})|
%\end{eqnarray}
Now, with respect to the inner product of spin-network functions, we can immediately gives the identity
\begin{equation}\label{id1}
\mathbb{I}^{\mathcal{H}_{\gamma}}:=\sum_{\vec{N}_{e}}(\prod_{e\in\gamma}\dim{\pi_{N_e}})\int_{\times_{v} \mathcal{P}_v}\prod_{v}d\vec{V}_{\vec{N}_{e_v}}|\gamma, \vec{N}_{e},\vec{\check{\mathcal{I}}}_v(\vec{V}_{e_v})\rangle\langle\gamma, \vec{N}_{e},\vec{\check{\mathcal{I}}}_v(\vec{V}_{e_v})|
\end{equation}
of the spin network function space $\mathcal{H}_{\gamma}$ spanned by the spin network states constructed on $\gamma$ whose edges are labelled by $\vec{N}$, where the sum over each $N_e$ take the range of non-negative integer.
Now, let us give the resolution of the identity of the space $\mathcal{H}_\gamma$ by using the coherent spin network state $|\gamma, \vec{\mathbb{H}}^o_{e}\rangle$,
which corresponds to the function
\begin{equation}
\tilde{\Psi}_{\gamma,\vec{\mathbb{H}}^o_{e}}(\vec{h}_{e}):= \sum_{\vec{N}_{e}}\left(\prod_{e}\dim(\pi_{N_{e}}) \exp(-\frac{(N_{e}-N_{e}^o)^2}{2\sigma^0_{e}})e^{-\mathbf{i}\xi^o_{e}N_{e}}\right) \cdot\Psi_{\gamma,\vec{N}_{e},\vec{\check{\mathcal{I}}}_{v}(\vec{V}_{e_v})}(\vec{h}_{e})
\end{equation}
with $N_{e}$ taking value in the set of non-negative integers while $N_{e}^o$ taking values in the whole real number set.
\\ \textbf{Theorem:} The identity of the quantum state space $\mathcal{H}_\gamma$ can be resolved by the coherent spin network state as
\begin{equation}\label{id2}
\mathbb{I}_{\mathcal{H}_\gamma}=\int_{\mathbb{R}^{\times|E(\gamma)|}}\prod_{e\in\gamma}dN^o_e\mathbf{C}_{\!\vec{N}^o} \int_{\times_{v} \mathcal{P}_v}\prod_{v}d\vec{V}_{e_v}\int_{\times_{e}S_e^1}\prod_{e\in\gamma}d\xi^o_{e}|\gamma, \vec{\mathbb{H}}^o_{e}\rangle\langle \gamma, \vec{\mathbb{H}}^o_{e}|, %\quad \sigma^0_e\sim \sqrt{N_e},
\end{equation}
where $dN^o_e$ is the Lebesgue measure on $\mathbb{R}$, $|E(\gamma)|$ represents the number of the edges of $\gamma$, and $d\vec{V}_{e_v}:=\prod_{e: b(e)=v}dV_{e}$ with $\int_{\times_{v} \mathcal{P}_v}\prod_{v}d\vec{V}_{e_v}=1$ and $d\vec{V}_{\vec{N}_{e_v}}=\prod_{e\in\gamma}\left(\dim(\pi_{N_e})\right)^2\cdot d\vec{V}_{e_v}$ by their definitions (note that we also re-orient the edges
 to be outgoing at each $v$ for the convenience of specific calculation here). Besides, the measure factor $\mathbf{C}_{\vec{N}^o_e}$ in Eq.\eqref{id2} is a function of $\vec{N}^o_e$ which satisfies
\begin{equation}\label{CN}
%\frac{1}{\prod_{e\in\gamma}\dim{(\pi_{N_e})}}=\sum_{\vec{N}^o_e} \frac{\mathbf{C}_{\!\vec{N}^o}}{\prod_{e\in\gamma}\dim{(\pi_{N^o_e})}} \prod_{e\in \gamma}\exp(-\frac{(N_{e}-N_{e}^o)^2}{\sigma^0_{e}}),
\prod_{e\in \gamma}\dim(\pi_{N_{e}}) =\int_{\mathbb{R}^{\times|E(\gamma)|}}\prod_{e\in\gamma}dN^o_e\mathbf{C}_{\vec{N}^o} \prod_{e\in \gamma}\exp(-\frac{(N_{e}-N_{e}^o)^2}{\sigma^0_{e}}),
\end{equation}
 the solution of this equation is given in Appendix.
\\
\\
\textbf{Proof:} Notice that we have the inner product
 \begin{equation}
 \langle\gamma, \vec{N}''_{e},\vec{\check{\mathcal{I}}}''_v(\vec{V}''_{e_v})|\gamma, \vec{N}'_{e},\vec{\check{\mathcal{I}}}'_v(\vec{V}'_{e_v})\rangle= \delta_{(\vec{N}',\vec{N}'')} \frac{1 }{\prod_{e\in\gamma}\dim{(\pi_{N'_e})}} \prod_{v}\langle \vec{\check{\mathcal{I}}}''_v(\vec{N}'_{e_v},\vec{V}''_{e_v})| \vec{\check{\mathcal{I}}}'_v(\vec{N}'_{e_v},\vec{V}'_{e_v})\rangle
 \end{equation}
 between the spin network functions, then we can verify that Eq.\eqref{id2} is the resolution of the identity of the quantum space $\mathcal{H}_\gamma$ by
\begin{eqnarray}
&&\int_{\mathbb{R}^{\times|E(\gamma)|}}\prod_{e\in\gamma}dN^o_e\mathbf{C}_{\!\vec{N}^o} \int_{\times_{v}\mathcal{P}_v}\prod_{v\in \gamma}d\vec{V}_{e_v}\int_{\times_{e}S_e^1}\prod_{e\in\gamma}d\xi^o_{e}\langle\gamma, \vec{N}''_{e},\vec{\check{\mathcal{I}}}''_v(\vec{V}''_{e_v})|\gamma, \vec{\mathbb{H}}^o_{e}\rangle\langle \gamma, \vec{\mathbb{H}}^o_{e}|\gamma, \vec{N}'_{e},\vec{\check{\mathcal{I}}}'_v(\vec{V}'_{e_v})\rangle\\\nonumber
&=&\int_{\mathbb{R}^{\times|E(\gamma)|}}\prod_{e\in\gamma}dN^o_e\mathbf{C}_{\!\vec{N}^o} \int_{\times_{v}\mathcal{P}_v}\prod_{v\in \gamma}d\vec{V}_{e_v}\int_{\times_{e}S_e^1}\prod_{e\in\gamma}d\xi^o_{e} \prod_{e}\exp(-\frac{(N''_{e}-N_{e}^o)^2}{2\sigma^0_{e}})e^{-\mathbf{i}\xi^o_{e}N''_{e}}\\\nonumber
&&\cdot \exp(-\frac{(N'_{e}-N_{e}^o)^2}{2\sigma^0_{e}})e^{\mathbf{i}\xi^o_{e}N'_{e}}\prod_{v}\langle \vec{\check{\mathcal{I}}}''_v(\vec{N}''_{e_v},\vec{V}''_{e_v})|\vec{\check{\mathcal{I}}}_v(\vec{N}''_{e_v},\vec{V}_{e_v})\rangle \langle \vec{\check{\mathcal{I}}}_v(\vec{N}'_{e_v},\vec{V}_{e_v})| \vec{\check{\mathcal{I}}}'_v(\vec{N}'_{e_v},\vec{V}'_{e_v})\rangle\\\nonumber
  &=& \delta_{(\vec{N}',\vec{N}'')} \int_{\mathbb{R}^{\times|E(\gamma)|}}\prod_{e\in\gamma}dN^o_e\mathbf{C}_{\!\vec{N}^o} \prod_{e}\exp(-\frac{(N'_{e}-N_{e}^o)^2}{\sigma^0_{e}})\\\nonumber
 && \cdot\int_{\times_{v}\mathcal{P}_v}\prod_{v\in \gamma}d\vec{V}_{e_v} \prod_{v}\langle \vec{\check{\mathcal{I}}}''_v(\vec{N}'_{e_v},\vec{V}''_{e_v})|\vec{\check{\mathcal{I}}}_v(\vec{N}'_{e_v},\vec{V}_{e_v})\rangle \langle \vec{\check{\mathcal{I}}}_v(\vec{N}'_{e_v},\vec{V}_{e_v})| \vec{\check{\mathcal{I}}}'_v(\vec{N}'_{e_v},\vec{V}'_{e_v})\rangle \\\nonumber
&=& \delta_{(\vec{N}',\vec{N}'')} \frac{1 }{(\prod_{e\in\gamma}\dim{(\pi_{N'_e})})^2} \int_{\mathbb{R}^{\times|E(\gamma)|}}\prod_{e\in\gamma}dN^o_e \mathbf{C}_{\!\vec{N}^o}\prod_{e}\exp(-\frac{(N'_{e}-N_{e}^o)^2}{\sigma^0_{e}}) \prod_{v}\langle \vec{\check{\mathcal{I}}}''_v(\vec{N}'_{e_v},\vec{V}''_{e_v})| \vec{\check{\mathcal{I}}}'_v(\vec{N}'_{e_v},\vec{V}'_{e_v})\rangle \\\nonumber
&=& \delta_{(\vec{N}',\vec{N}'')} \frac{1 }{\prod_{e\in\gamma}\dim{(\pi_{N'_e})}} \prod_{v}\langle \vec{\check{\mathcal{I}}}''_v(\vec{N}'_{e_v},\vec{V}''_{e_v})| \vec{\check{\mathcal{I}}}'_v(\vec{N}'_{e_v},\vec{V}'_{e_v})\rangle,
\end{eqnarray}
where we used \eqref{id222} and the definition of $d\vec{V}_{e_v}$
 %$\prod_{v\in\gamma}\prod_{e_v}\dim{(\pi_{N'_{e_v}})}=(\prod_{e\in\gamma}\dim{(\pi_{N'_e})})^2$
  in the third equal. This finish the proof. \\$\square$\\
  This resolution of identity for gauge invariant case can be given by projecting the identity \eqref{id2} to the gauge invariant cylindrical function space $\underline{\mathcal{H}}_\gamma$ constructed on $\gamma$, that is,
\begin{equation}\label{id3}
\mathbb{I}_{\underline{\mathcal{H}}_\gamma}=\int_{\mathbb{R}^{\times|E(\gamma)|}}\prod_{e\in\gamma}dN^o_e\mathbf{C}_{\!\vec{N}^o} \int_{\times_{v} \mathcal{P}_v}\prod_{v}d\vec{V}^o_{e_v}\int_{\times_{e}S_e^1}\prod_{e\in\gamma}d\xi^o_{e} \mathbb{P}_{\text{inv}}|\gamma, \vec{\mathbb{H}}^o_{e}\rangle\langle \gamma, \vec{\mathbb{H}}^o_{e}|\mathbb{P}_{\text{inv}},
\end{equation}
where $\mathbb{P}_{\text{inv}}$ is the projection operator to space $\underline{\mathcal{H}}_\gamma$ and it is realized by taking the group averaging at each of the vertice of the state $|\gamma, \vec{\mathbb{H}}^o_{e}\rangle$. The result of the group averaging is just the gauge invariant superposition type coherent state $\underline{|\gamma, \vec{\mathbb{H}}^o_{e}}\rangle$ which corresponds to the function \eqref{gispcs}. Hence, the identity \eqref{id3} can be simplified as
\begin{equation}\label{id4}
\mathbb{I}_{\underline{\mathcal{H}}_\gamma}=\int_{\mathbb{R}^{\times|E(\gamma)|}}\prod_{e\in\gamma}dN^o_e\mathbf{C}_{\!\vec{N}^o} \int_{\times_{v} \mathcal{P}_v}\prod_{v}d\vec{V}_{e_v}\int_{\times_{e}S_e^1}\prod_{e\in\gamma}d\xi^o_{e} \underline{|\gamma, \vec{\mathbb{H}}^o_{e}}\rangle\langle\underline{ \gamma, \vec{\mathbb{H}}^o_{e}|}.
\end{equation}
\section{Peakedness property for a simple example}

 The obvious advantage of these super-position type coherent states is that they take the Gaussian super-position formulation over the spin net-work states labelled by simple coherent intertwiners. This fact indicates that they are peaked at their labels and this peakedness is Gaussian damped. However, this Gaussian super-position is not complete, because the sum in Eqs. \eqref{spcs} and \eqref{gispcs} takes over the non-negative integer $N$ rather that whole integers. This problem can be avoided in large $N^o$ limit as follows. Note that the factor $\exp(-\frac{(N-N^o)^2}{2\sigma^0_{e}})$ in Eqs. \eqref{spcs} and \eqref{gispcs} tends to zero with $N^o\to\infty$ and $N<0$ so that the contribution of the terms with $N<0$ to the the sum over $N$ in Eqs. \eqref{spcs} and \eqref{gispcs} vanishes in large $N^o$ limit. Hence the Gaussian super-position in Eqs. \eqref{spcs} and \eqref{gispcs} is complete in large $N^o$ limit and we can expect a well localization properties of this coherent states in this case.

  In order to show this property more clearly, we consider a rather simple example: the coherent loop. This example allows us to discuss the importance of the appropriate choice of heat-kernel time $t_{e}$.
 When the graph is given by a loop $\gamma$, the dependence of the gauge invariant super-position type coherent state on the bi-vectors $\vec{V}$
  drops out and the state is simply labeled by a couple of variables $(\xi^o,N^o)$, then we find
 \begin{equation}\label{loop}
\tilde{\Psi}_{\gamma,(\xi^o,N^o)}(h)= \sum_{N} \exp(-\frac{(N-N^o)^2}{2\sigma^0})e^{-\mathbf{i}\xi^oN} \cdot \text{tr}^{(N)}(h).
\end{equation}
Now we compute the expectation value of the area operator $\hat{A}$ for a (D-1)-dimensional surface that is punctured once by the loop. As well known, we have
\begin{equation}
  \hat{A}\text{tr}^{(N)}(h)=16\pi\beta(l_p^{(D+1)})^{D-1}\sqrt{N(N+D-1)}\text{tr}^{(N)}(h).
\end{equation}
Therefore
 \begin{equation}
\hat{A}\Psi_{\gamma,(\xi^o,N^o)}(h)= 16\pi\beta(l_p^{(D+1)})^{D-1}\sum_{N} \exp(-\frac{(N-N^o)^2}{2\sigma^0})e^{-\mathbf{i}\xi^oN}\sqrt{N(N+D-1)} \cdot \text{tr}^{(N)}(h).
\end{equation}
In the limit of large $N^o$, the expectation value of the (D-1)-area operator is easily computed
\begin{equation}
\langle\hat{A}\rangle=\frac{\left(\Psi_{\gamma,(\xi^o,N^o)}(h) ,\hat{A}\Psi_{\gamma,(\xi^o,N^o)}(h)\right)}{\left(\Psi_{\gamma,(\xi^o,N^o)}(h), \Psi_{\gamma,(\xi^o,N^o)}(h)\right)}=16\pi\beta(l_p^{(D+1)})^{D-1}\sqrt{N^o(N^o+D-1)}
\end{equation}
and confirms the interpretation of $N^o$ as the quantity that prescribes the expectation value of the (D-1)-area, where $(\Psi_1,\Psi_2)$ denotes the inner product of the states $\Psi_1$ and $\Psi_2$.

Now we consider another operator acting on the Hilbert space $\mathcal{H}_\gamma$: the Wilson loop operator $W_\gamma$. Recall that it acts on basis states as
\begin{equation}
\hat{W}_\gamma\text{tr}^{(N)}(h)=\text{tr}^{(1)}(h)\text{tr}^{(N)}(h) =\text{tr}^{(N+1)}(h)+\text{tr}^{(N-1)}(h)+\text{tr}^{(\text{not simple})}(h).
\end{equation}
where $\text{tr}^{(\text{not simple})}(h)$ denote the terms of trace which are not belonging to simple representations. As a result, we find for large $N^o$, we have
\begin{eqnarray}
\langle\hat{W}_\gamma\rangle&=&\frac{\sum_{N}\exp(-\frac{(N-N^o)^2+(N+1-N^o)^2}{2\sigma^0}) e^{\mathbf{i}\xi^o}+\sum_{N}\exp(-\frac{(N-N^o)^2+(N+1-N^o)^2}{2\sigma^0}) e^{-\mathbf{i}\xi^o}}{\sum_{N}\exp(-\frac{(N-N^o)^2}{\sigma^0}) }\\\nonumber
&=&2\cos\xi^o\sum_{N}\exp(-\frac{(N-N^o)^2+(N+1-N^o)^2}{2\sigma^0})/(\sum_{N}\exp(-\frac{(N-N^o)^2}{\sigma^0})\\\nonumber
&=&2\cos\xi^oe^{-\frac{1}{4\sigma^0}}.
\end{eqnarray}
Therefore, in the limit $\sigma^0\rightarrow\infty$ compatible with $N^o$ large, the $\xi^o$ identifies where the Wilson loop is peaked on.
Similarly, we can compute the dispersions of the area operator and of the Wilson
loop. We find
\begin{equation}
\Delta\langle\hat{A}\rangle=\sqrt{\langle\hat{A}^2\rangle-\langle\hat{A}\rangle^2}=8\pi\beta(l_p^{(D+1)})^{D-1}{\sqrt{2\sigma^0}},
\end{equation}
\begin{equation}
\Delta\langle\hat{W}_\gamma\rangle=\sqrt{\langle\hat{W}_\gamma^2\rangle-\langle\hat{W}_\gamma\rangle^2} =\sqrt{2\cos(2\xi^o)e^{-\frac{1}{\sigma^0}}+3 -4\cos^2\xi^oe^{-\frac{1}{2\sigma^0}}}=\sqrt{1+\sin^2\xi^o\frac{2}{\sigma^0}},
\end{equation}
where we used that
\begin{equation}
\hat{W}_\gamma\hat{W}_\gamma\text{tr}^{(N)}(h)=\text{tr}^{(1)}(h)\text{tr}^{(1)}(h)\text{tr}^{(N)}(h) =\text{tr}^{(N+2)}(h)+3\text{tr}^{(N)}(h)+\text{tr}^{(N-2)}(h)+\text{tr}^{(\text{not simple})}(h).
\end{equation}
and taken the limit $\sigma^0\rightarrow\infty$ compatible with $N^o$ large.

As we can see, the results of the dispersion of the expectation value of Wilson loop operator is not proportional to $\frac{1}{\sqrt{\sigma^0}}$, which does not repeat the result of the similar calculation in the (1+3)-dimensional $SU(2)$ LQG \cite{Bianchi_2010}. In fact, different with the $SU(2)$ Wilson loop, we can check that the $ SO(D+1)$ Wilson loop given by
\begin{equation}
W_\gamma=\text{tr}(h_\gamma)=2\cos(\xi^o)\text{tr}(e^{\bar{\xi}^\mu\bar{\tau}_\mu})
\end{equation}
relies on a group element $e^{\bar{\xi}^\mu\bar{\tau}_\mu}\in SO(D-1)\subset SO(D+1)$ which may related to the gauge degrees of freedom with respect to simplicity constraint, instead of only relies on the $\xi^o$ which capture the degrees of freedom of extrinsic curvature. Also, it is easy to see that the $ SO(D+1)$ Wilson loop operator does not commute with the simplicity constraint by
\begin{equation}
(\hat{W}_\gamma\mathbb{P}_{\text{s.}}-\mathbb{P}_{\text{s.}}\hat{W}_\gamma)\text{tr}^{(N)}(h)=\text{tr}^{(1)}(h)\text{tr}^{(N)}(h) -\mathbb{P}_{\text{s.}}\text{tr}^{(1)}(h)\text{tr}^{(N)}(h)  =\text{tr}^{(\text{not simple})}(h),
\end{equation}
where $\mathbb{P}_{\text{s.}}$ is a projector which projects the state to the space composed by all of the spin network functions whose edges are labelled by simple representations.
Hence the Wilson loop operator is not a physical kinematic observable and it contains the gauge degrees of freedom corresponding to simplicity constraint. In order to avoid this problem, we can define the physical Wilson loop operator as $\mathbb{P}_{\text{s.}}\hat{W}_\gamma\mathbb{P}_{\text{s.}}$. It acts on the spin network state on a loop as
 \begin{equation}
\mathbb{P}_{\text{s.}}\hat{W}_\gamma\mathbb{P}_{\text{s.}}\text{tr}^{(N)}(h)=\mathbb{P}_{\text{s.}}\text{tr}^{(1)}(h)\text{tr}^{(N)}(h) =\text{tr}^{(N+1)}(h)+\text{tr}^{(N-1)}(h).
\end{equation}
 It is easy to see that $\mathbb{P}_{\text{s.}}\hat{W}_\gamma\mathbb{P}_{\text{s.}}$ is commutative with the quantum simplicity constraint. Its expectation value and the corresponding dispersion is given by
\begin{eqnarray}
\langle\mathbb{P}_{\text{s.}}\hat{W}_\gamma\mathbb{P}_{\text{s.}}\rangle&=&2\cos\xi^oe^{-\frac{1}{4\sigma^0}},
\end{eqnarray}
\begin{equation}
\Delta\langle\mathbb{P}_{\text{s.}}\hat{W}_\gamma\mathbb{P}_{\text{s.}}\rangle =\sqrt{\langle(\mathbb{P}_{\text{s.}}\hat{W}_\gamma\mathbb{P}_{\text{s.}})^2\rangle -\langle\mathbb{P}_{\text{s.}}\hat{W}_\gamma\mathbb{P}_{\text{s.}}\rangle^2} =\sqrt{2\cos(2\xi^o)e^{-\frac{1}{\sigma^0}}+2 -4\cos^2\xi^oe^{-\frac{1}{2\sigma^0}}}=\sqrt{\frac{2}{\sigma^0}\sin^2\xi^o}
\end{equation}
where we taken the limit $\sigma^0\rightarrow\infty$ compatible with $N^o$ large again. Then we can conclude
\begin{equation}
\Delta\langle\hat{A}\rangle\cdot\Delta\langle\mathbb{P}_{\text{s.}}\hat{W}_\gamma\mathbb{P}_{\text{s.}}\rangle \propto \sin\xi^o,
\end{equation}
which minimalize the uncertainty relation given by the non-vanishing Poisson bracket $\{\cos(\xi^o), N^o\}\propto \sin(\xi^o)$ based on Eq. \eqref{xiN}.

Now notice that, as the area and the Wilson loop are non-commuting operators, we cannot
make both their dispersions vanish at the same time. Large $\sigma^0$ means that
the state is sharply peaked on the holonomy, while small $\sigma^0$ means that
the state is sharply peaked on the (D-1)-area. A good requirement of semiclassicality is that
the relative dispersions of both operators vanish in the large $N^o$ limit. Let us assume $\sigma^0\sim (N^o)^{-k}$, using the results
derived above, we find the following behavior for relative dispersions:
\begin{equation}\label{bound2}
  \frac{\Delta\langle\hat{A}\rangle}{\langle\hat{A}\rangle}\sim(N^o)^{-\frac{k+2}{2}},\quad \frac{\Delta \langle\mathbb{P}_{\text{s.}}\hat{W}_\gamma\mathbb{P}_{\text{s.}}\rangle} {\langle\mathbb{P}_{\text{s.}}\hat{W}_\gamma\mathbb{P}_{\text{s.}}\rangle}\sim(N^o)^{\frac{k}{2}}.
\end{equation}
The first requires $k > -2$ and the second $k < 0$. Moreover, one hopes that the coherent state Eq.\eqref{loop} can be given by a heat kernel coherent state with taking large $\eta$ limit. Then, due to the relation \eqref{etat}, the limit ``large $\eta$ and large $N^o$'' can be attained only if $t=\frac{1}{2\sigma^0}$ scales with $N^o$ as
\begin{equation}\label{bound1}
t\sim (N^o)^k, \quad k>-1.
\end{equation} 
Taking into account the three bounds \eqref{bound2}
\eqref{bound1} we find that the coherent loop behaves semiclassically when the heat-kernel time $t$ scales as $(N^o)^k$ with $-1 < k < 0$. For instance, the choice $t = \frac{1}{\sqrt{N^o}}$ guarantees the semi-classicality of the state.
\section{ Conclusion}
This paper proposes the superposition type coherent state in all dimensional LQG. Comparing it with the heat-kernel coherent state for $SO(D+1)$, it is shown that these two types of coherent states are consistent in the large $\eta$ limit. Moreover, it turns out that the superposition type coherent states have well peakedness property in large $N^o$ limit and serve a resolution of identity in the Hilbert space where it belongs. Here, it is noted that, for small and negative valued $N^o$, the superposition type coherent states are not well-peaked because they do not minimalized uncertainty, but these states are still sufficient to construct the resolution of the identity.
Finally, considering the one loop graph as an example, we show that superposition type coherent states minimalize the quantum uncertainty between the area operator and the physical Wilson loop operator, where the uncertainty of these operators also serve a limitation on the heat kernel time $t$ to guarantee the semi-classicality of the heat kernel coherent state.

Comparing with the complicated heat-kernel coherent states, the superposition type $SO(D+1)$ coherent states serves us a practical tool to explore the semi-classicality of $(1+D)$-dimensional LQG. However, this tool is only valid in the large $N^o$ limit, which limits its application scope. It will be our further research to overcome this defect.
\section*{Acknowledgments}
This work is supported by the National Natural Science Foundation of China (NSFC) with Grants No. 12047519, No. 11775082, No. 11875006 and No. 11961131013. C. Z. acknowledges the support by the Polish Narodowe Centrum Nauki, Grant No. 2018/30/Q/ST2/00811

\appendix

\section{The measure factor in the resolution of identity}
The solution of Eq.\eqref{CN} can be given by $\mathbf{C}_{\!\vec{N}^o}=\prod_{e\in \gamma}f(N_e^o)$ with each of $f(N_e^o)$ satisfying
\begin{equation}\label{simplify}
\dim(\pi_N)=\frac{(D+N-2)!(2N+D-1)}{(D-1)!N!}=\int_{\mathbb{R}}dN^of(N^o)\exp(-\frac{(N-N^o)^2}{\sigma^0}).
\end{equation}
Notice that $\frac{(D+N-2)!(2N+D-1)}{(D-1)!N!}=\frac{(N+D-2)(N+D-3)\cdot...\cdot(N+1)(2N+D-1)}{(D-1)!}$ is a polynomial of $N$.Then, solving $f(N^o)$ satisfying Eq.\eqref{simplify} is equivalent to solve the equation
\begin{equation}\label{fny}
x^n=\int_{\mathbb{R}}dyf_n(y)\exp(-\frac{(x-y)^2}{\sigma^0}), \ \  x\in\mathbb{Z}_+,\ \ n\in \mathbb{Z}_+.
\end{equation}
 This can be achieved as follows.
By defining
\begin{equation}
\text{F}_n=\int_{\mathbb{R}}dyy^n\exp(-\frac{y^2}{\sigma^0}),
\end{equation}
and using $y^n=\sum_{\ell=1}^n\text{C}^n_{\ell}(y-x) ^\ell x^{(n-\ell)}$, we have
\begin{equation}
n\text{\ is\ odd}:\quad\text{F}_n=\int_{\mathbb{R}}dyy^n\exp(-\frac{(x-y)^2}{\sigma^0})=x\text{C}^n_{n-1}\text{F}_{n-1}+x^3\text{C}^n_{n-3}\text{F}_{n-3}+... +x^n\text{C}^n_{0}\text{F}_{0},
\end{equation}
\begin{equation}
n\text{\ is\ even}:\quad\text{F}_n=\int_{\mathbb{R}}dyy^n\exp(-\frac{(x-y)^2}{\sigma^0})=\text{C}^n_{n}\text{F}_{n}+x^2 \text{C}^n_{n-2}\text{F}_{n-2}+... +x^n\text{C}^n_{0}\text{F}_{0}.
\end{equation}
Then, by setting $n=0, n=1, n=2,...$ step by step, we have
\begin{eqnarray}
  f_0(y) &=& \frac{1}{\text{C}^0_{0}\text{F}_{0}}, \\
 f_1(y) &=& \frac{y}{\text{C}^1_{0}\text{F}_{0}}, \\
f_2(y) &=& \frac{y^2-\text{C}^2_2\text{F}_2f_0(y)}{\text{C}^2_0\text{F}_0},\\
 f_3(y) &=& \frac{y^3-\text{C}^3_2\text{F}_2f_1(y)}{\text{C}^3_0\text{F}_0},\\
 f_4(y) &=& \frac{y^4-\text{C}^4_2\text{F}_2f_2(y)-\text{C}^4_4\text{F}_4f_0(y)}{\text{C}^4_0\text{F}_0},\\
 f_5(y) &=& \frac{y^5-\text{C}^5_2\text{F}_2f_3(y)-\text{C}^5_4\text{F}_4f_1(y)}{\text{C}^5_0\text{F}_0},\\
 &&...\\\nonumber
 f_n(y)&=&\frac{y^n-\text{C}^n_2\text{F}_2f_{n-2}(y)-...-\text{C}^n_{n-1}\text{F}_{n-1}f_1(y)}{\text{C}^n_0\text{F}_0},\quad n\text{\ is\ odd},\\
f_n(y)&=&\frac{y^n-\text{C}^n_2\text{F}_2f_{n-2}(y)-...-\text{C}^n_{n-2}\text{F}_{n-2}f_{2}(y)-\text{C}^n_{n}\text{F}_{n}f_0(y)}{\text{C}^n_0\text{F}_0},\quad n\text{\ is\ even}.
\end{eqnarray}
Finally, the solution $f_n(y)$ of Eq. \eqref{fny} can be got by following this algorithm.
\bibliographystyle{unsrt}

\bibliography{ref}
\end{document}